\documentclass{article}

\usepackage{arxiv}

\usepackage{amsthm}
\usepackage[utf8]{inputenc} % allow utf-8 input
\usepackage[T1]{fontenc}    % use 8-bit T1 fonts
\usepackage{hyperref}       % hyperlinks
\usepackage{url}            % simple URL typesetting
\usepackage{booktabs}       % professional-quality tables
\usepackage{amssymb}
\usepackage{amsbsy}
\usepackage{amsfonts}       % blackboard math symbols
\usepackage{nicefrac}       % compact symbols for 1/2, etc.
\usepackage{microtype}      % microtypography
\usepackage{lipsum}
\usepackage{amsmath}
\usepackage[utf8]{inputenc}

\newcommand{\ee}[1]{\emph{#1}}

\usepackage{listings}
\usepackage{xcolor}
 
\definecolor{codegreen}{rgb}{0,0.6,0}
\definecolor{codegray}{rgb}{0.5,0.5,0.5}
\definecolor{codepurple}{rgb}{0.58,0,0.82}
\definecolor{backcolour}{rgb}{0.95,0.95,0.92}
 
\lstdefinestyle{mystyle}{
    backgroundcolor=\color{backcolour},   
    commentstyle=\color{codegreen},
    keywordstyle=\color{magenta},
    numberstyle=\tiny\color{codegray},
    stringstyle=\color{codepurple},
    basicstyle=\ttfamily\footnotesize,
    breakatwhitespace=false,         
    breaklines=true,                 
    captionpos=b,                    
    keepspaces=true,                 
    numbers=left,                    
    numbersep=5pt,                  
    showspaces=false,                
    showstringspaces=false,
    showtabs=false,                  
    tabsize=2
}
 
\usepackage{graphicx}

\title{scikit-mobility: a Python library for the analysis, generation and risk assessment of mobility data}

\author{Luca Pappalardo\\
ISTI-CNR, Italy \\
luca.pappalardo@isti.cnr.it
   \And 
   Filippo Simini\\
   University of Bristol, UK \\ Argonne National Lab, US \\
   f.Simini@bristol.ac.uk \\
   \And Gianni Barlacchi *\\
   FBK, Italy\\ Amazon Alexa, Germany \\
   barlacchi@fbk.eu
   \And Roberto Pellungrini\\
   University of Pisa, Italy\\
   roberto.pellungrini@gmail.com}

\begin{document}
\maketitle

\begin{abstract} 
The last decade has witnessed the emergence of massive mobility data sets, such as tracks generated by GPS devices, call detail records, and geo-tagged posts from social media platforms. These data sets have fostered a vast scientific production on various applications of mobility analysis, ranging from computational epidemiology to urban planning and transportation engineering. 
A strand of literature addresses data cleaning issues related to raw spatiotemporal trajectories, while the second line of research focuses on discovering the statistical ``laws'' that govern human movements. 
A significant effort has also been put on designing algorithms to generate synthetic trajectories able to reproduce, realistically, the laws of human mobility.
Last but not least, a line of research addresses the crucial problem of privacy, proposing techniques to perform the re-identification of individuals in a database. 
A view on state of the art cannot avoid noticing that there is no statistical software that can support scientists and practitioners with all the aspects mentioned above of mobility data analysis.
In this paper, we propose scikit-mobility, a Python library that has the ambition of providing an environment to reproduce existing research, analyze mobility data, and simulate human mobility habits. scikit-mobility is efficient and easy to use as it extends pandas, a popular Python library for data analysis. Moreover, scikit-mobility provides the user with many functionalities, from visualizing trajectories to generating synthetic data, from analyzing statistical patterns to assessing the privacy risk related to the analysis of mobility data sets.
\end{abstract}

% keywords can be removed

\vfill
\textbf{*} Work done prior joining Amazon

\theoremstyle{definition}
\newtheorem{definition}{Definition}[section]

\vfill
\section{Introduction}
The last decade has witnessed the emergence of massive datasets of digital traces that portray human movements at an unprecedented scale and detail. 
Examples include tracks generated by GPS devices embedded in personal smartphones \cite{zheng2008geolife}, private vehicles \cite{pappalardo2013understanding} or boats \cite{fernandez2017maritime}; call detail records produced as a by-product of the communication between cellular phones and the mobile phone network \cite{gonzalez08understanding, barlacchi2015multi}; geotagged posts from the most disparate social media platforms \cite{noulas2012tale}; even traces describing the sports activity of amateurs or professional athletes \cite{rossi2017effective}. 
The availability of digital mobility data has attracted enormous interests from scientists of diverse disciplines, fueling advances in several applications, from computational health \cite{tizzoni2012realtime, barlacchi2017you} to the estimation of air pollution \cite{nyhan2018quantifying, bohm2021quantifying}, from the design of recommender systems \cite{wang2011human} to the optimization of mobile and wireless networks \cite{karamshuk2011human, tomasini2017effect}, from transportation engineering and urban planning \cite{zhao2016urban} to the estimation of migratory flows \cite{simini2012universal, ahmed2016multi} and people's place of residence \cite{pappalardo2021evaluation, vanhoof2020performance}, from the well-being status of municipalities, regions and countries \cite{pappalardo2016analytical, voukelatou2020measuring} to the prediction of traffic and future displacements \cite{zhang2017deep, rossi2019modelling}.

It is hence not surprising that the last decade has also witnessed a vast scientific production on various aspects of human mobility \cite{luca2020deep, wang2019urban, blondel2015survey, barbosa2018human}.
The first strand of literature addresses data preprocessing issues related to mobility data, such as how to extract meaningful locations from raw spatiotemporal trajectories, how to filter, reconstruct, compress and segment them, or how to cluster and classify them \cite{zheng2015trajectory}. 
As a result, in the literature, there is a vast repertoire of techniques that allow scientists and professionals to improve the quality of their mobility data. 
 
The second line of research focuses on discovering the statistical laws that govern human mobility. These studies document that, far from being random, human mobility is characterized by predictable patterns, such as a stunning heterogeneity of human travel patterns \cite{gonzalez08understanding}; a strong tendency to routine and a high degree of predictability of individuals' future whereabouts \cite{song2010limits}; the presence of the returners and explorers dichotomy \cite{pappalardo2015returners}; a conservative quantity in the number of locations actively visited by individuals \cite{alessandretti2018evidence}, and more \cite{barbosa2018human, luca2020deep}. 
These quantifiable patterns are universal across different territories and data sources and are usually referred to as the ``laws'' of human mobility.

The third strand of literature focuses on designing generative algorithms, i.e., models that can generate synthetic trajectories able to reproduce, realistically, the laws of human mobility. 
A class of algorithms aims to reproduce spatial properties of mobility \cite{song2010modelling, pappalardo2016human}; 
another one focuses on the accurate representation of the time-varying behavior of individuals \cite{barbosa2015effect, alessandretti2018evidence}. 
More recently, some approaches rely on machine learning to propose generative algorithms that are realistic with respect to both spatial and temporal properties of human mobility \cite{pappalardo2018data, jiang2016timegeo, luca2020deep}. 
Although the generation of realistic trajectories is a complex and still open problem, the existing algorithms act as baselines for the evaluation of new approaches.

Finally, a line of research addresses the crucial problem of privacy: people's movements might reveal confidential personal information or allow the re-identification of individuals in a database, creating serious privacy risks \cite{demontoye2013unique, fiore2020privacy}. 
Since 2018, the EU General Data Protection Regulation (GDPR) explicitly imposes on data controllers an assessment of the impact of data protection for the riskiest data analyses. Driven by these sensitive issues, in recent years researchers have developed algorithms, methodologies, and frameworks to estimate and mitigate the individual privacy risks associated with the analysis of digital data in general \cite{monreale2014privacy} and mobility records in particular \cite{pellungrini2017data, pellungrini2020modeling, demontoye2013unique, demontoye2018privacy}.

Despite the increasing importance of mobility analysis for many scientific and industrial domains, there is no statistical software that can support scientists and practitioners with all the aspects of mobility analysis mentioned above (Section \ref{sec:related}).

To fill this gap, we propose \emph{scikit-mobility}, a python library that has the ambition of providing scientists and practitioners with an environment to reproduce existing research and perform analysis of mobility data. In particular, the library allows the user to: 

\begin{enumerate}

\item load and represent mobility data, both at the individual and the collective level, through easy-to-use data structures (\texttt{TrajDataFrame} and \texttt{FlowDataFrame})  based on the standard python libraries \emph{numpy} \cite{numpy}, \emph{pandas} \cite{pandas} and \emph{geopandas} \cite{geopandas} (Section \ref{sec:data_structures}), as well as to visualize trajectories and flows on interactive maps based on the python libraries \emph{folium} \cite{folium} and \emph{matplotlib} \cite{matplotlib} (Section \ref{sec:plotting});

    \item clean and preprocess mobility data using state-of-the-art techniques, such as trajectory clustering, compression, segmentation, and filtering. The library also provides the user with a way to track all the operations performed on the original data (Section \ref{sec:preprocessing});

\item analyze mobility data by using the main measures characterizing mobility patterns both at the individual and at the collective level (Section \ref{sec:measures}), such as the computation of travel and characteristic distances, object and location entropies, location frequencies, waiting times, origin-destination matrices, and more;

\item run the most popular mechanistic generative models to simulate individual mobility, such as the Exploration and Preferential Return model (EPR) and its variants (Section \ref{sec:individual_models}), and commuting and migratory flows, such as the Gravity Model and the Radiation Model (Section \ref{sec:collective_models});

\item estimate the privacy risk associated with the analysis of a given mobility dataset through the simulation of the re-identification risk associated with a vast repertoire of privacy attacks (Section \ref{sec:privacy}).
\end{enumerate}

Next-location prediction, i.e., predicting the next location(s) an individual will visit given their mobility history \cite{luca2020deep, wu2018location}, is a relevant mobility-related task not covered in the current version of \emph{scikit-mobility}. 
We plan to include location prediction algorithms in future versions of the library.

Note that, while \emph{scikit-mobility} has been conceived for human movement analysis and the privacy module makes sense for human mobility data only, most features can be applied to other types of mobility (e.g., boats, animal movements, boat trips).
\emph{scikit-mobility} is designed to deal with spatiotemporal trajectories and mobility flows and functions to deal with other types of mobility-related data, such as accelerometer data from wearable devices, are not currently covered in this library.

Clearly, the methods currently implemented have been chosen by the authors based mostly on their expertise and are by no means meant to be exhaustive. In future releases of the library, we plan to expand the range of methods and models.

\emph{scikit-mobility} is publicly available on GitHub at the following link: \url{https://scikit-mobility.github.io/scikit-mobility/}. 
Tutorials on how to use the library for mobility analysis is available at the following link: \url{https://github.com/scikit-mobility/tutorials}.
The documentation describing all the classes and functions of \emph{scikit-mobility} is available at \url{https://scikit-mobility.github.io/scikit-mobility/}.

\section{Data Structures} \label{sec:data_structures}

\emph{scikit-mobility} provides two data structures to deal with raw trajectories and flows between places. Both the data structures are an extension of the DataFrame implemented in the data analysis library \emph{pandas} \cite{pandas}. 
Thus, both \texttt{TrajDataFrame} and \texttt{FlowDataFrame} inherit all the functionalities provided by the DataFrame as well as all the efficient optimizations for reading and writing tabular data (e.g., mobility datasets).
This choice allows broad compatibility of \emph{scikit-mobility} with other python libraries and machine learning tools, such as \ee{scikit-learn}. 

Note that the current version of the library is designed to work with the latitude and longitude system (\texttt{epsg:4326}), the most used one in practical scenarios of mobility analysis.
Therefore, the Haversine formula is used by default when the library's functions compute distances. 
We plan to extend the library to deal with other reference systems, even user-defined ones. 
This extension would imply associating a custom distance function to a reference system.

\subsection{Trajectory}
Mobility data describe the movements of a set of objects during a period of observation. 
The objects may represent individuals \cite{gonzalez08understanding}, animals \cite{ramos2004levy}, private vehicles \cite{pappalardo2015returners}, boats \cite{fernandez2017maritime} and even players on a sports field \cite{rossi2017effective}.
Mobility data are generally collected in an automatic way as a by-product of human activity on electronic devices (e.g., mobile phones, GPS devices, social networking platforms, video cameras) and stored as \emph{trajectories}, a temporally ordered sequence of spatio-temporal points where an object stopped in or went through. 
In the literature of mobility analytics, a trajectory is often formally defined as follows \cite{zheng2014urban, zheng2015trajectory}: 
\begin{definition}[Trajectory]
The trajectory of an object $u$ is a temporally ordered sequence of tuples $T_u= \langle (l_1,t_1), (l_2,t_2), \dots, (l_n,t_n) \rangle$, where $l_i=(x_i,y_i)$ is a location, $x_i$ and $y_i$ are the coordinates of the location, and $t_i$ is the corresponding timestamp, with $t_i<t_j$ if $i<j$.
\end{definition}

\begin{figure}[]\centering
\includegraphics[width=0.6\columnwidth]{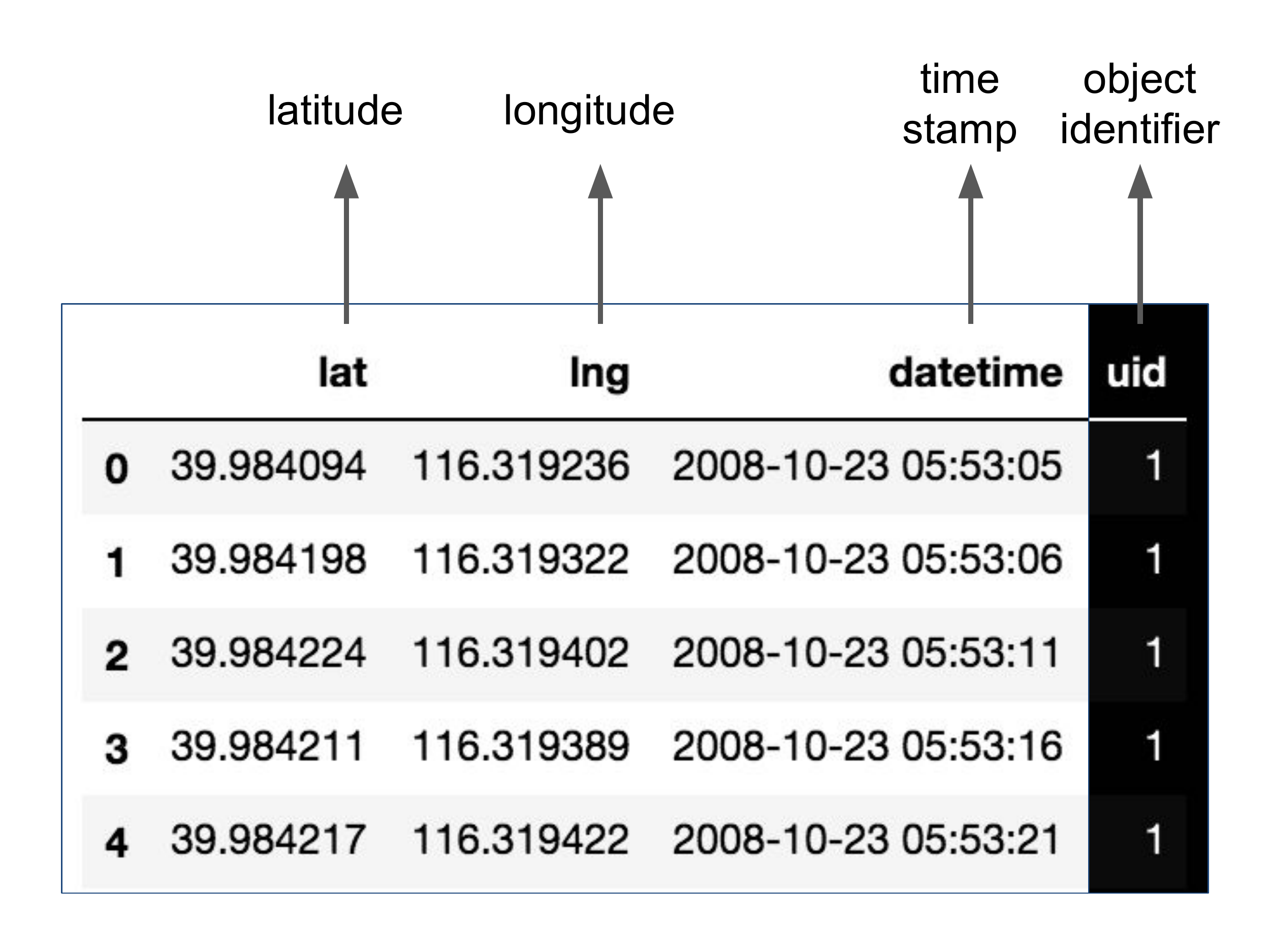}
\caption{Representation of a TrajDataFrame. Each row represents a point of an object's trajectory, described by three mandatory columns (\texttt{lat}, \texttt{lng}, \texttt{datetime}) and eventually by the column \texttt{uid} and \texttt{tid}, indicating the object associated with the point and the trajectory id, respectively.}
\label{fig:trajdataframe}
\end{figure}

In \emph{scikit-mobility}, a set of trajectories is described by a \texttt{TrajDataFrame} (Figure \ref{fig:trajdataframe}), an extension of the \emph{pandas} DataFrame that has specific columns names and data types. A row in the \texttt{TrajDataFrame} represents a point of the trajectory, described by three mandatory fields (aka columns): \texttt{latitude} (type: float), \texttt{longitude} (type: float) and \texttt{datetime} (type: datetime). 

Additionally, two optional columns can be specified. 
The first one is \texttt{uid}: it identifies the object associated with the point of the trajectory and can be of any type (string, int or float).
If \texttt{uid} is not present, \emph{scikit-mobility} assumes that the \texttt{TrajDataFrame} contains trajectories associated with a single moving object. 
The second one is \texttt{tid} (any type) and specifies the identifier of the trajectory to which the point belongs to.
If \texttt{tid} is not present, \emph{scikit-mobility} assumes that all the rows in the \texttt{TrajDataFrame} that are associated with a \texttt{uid} belong to the same trajectory. 
Note that, besides the mandatory columns, the user can add to a \texttt{TrajDataFrame} as many columns as they want since the data structures in \emph{scikit-mobility} inherit all the \emph{pandas} DataFrame functionalities.

Each \texttt{TrajDataFrame} also has two mandatory attributes: 

\begin{itemize}
\item \texttt{crs} (type: dictionary): indicates the coordinate reference system associated with the trajectories. By default it is \texttt{epsg:4326} (the latitude/longitude reference system);
\item \texttt{parameters} (type: dictionary): indicates the operations that have been applied to the \texttt{TrajDataFrame}. This attribute is a dictionary the key of which is the signature of the function applied (see Section \ref{sec:preprocessing} for more details).
\end{itemize}

\emph{scikit-mobility} provides functions to create a \texttt{TrajDataFrame} from mobility data stored in different formats (e.g., dictionaries, lists, \emph{pandas} DataFrames). 
To load a \texttt{TrajDataFrame} from a file, we first import the library.

\begin{verbatim}
Python> import skmob    
\end{verbatim}

Then, we use the method \texttt{from\_file} of the \texttt{TrajDataFrame} class to load the mobility data from the file path. 

\begin{verbatim}
Python> tdf = skmob.TrajDataFrame.from_file('geolife_sample.txt.gz')
\end{verbatim}

Note that the values corresponding to the \texttt{lat}, \texttt{lng}, and \texttt{datetime} columns must be necessarily float, float and datetime, respectively, otherwise the library raises an exception.\footnote{
The \texttt{TrajDataFrame} constructor forces the conversion of the values of the three mandatory columns to the preset types. Only if the conversion fails, it raises an exception. For example, the constructor can successfully convert string "39.1432" to float 39.1432, but it cannot convert (and hence raises an exception) string "39.2ui2" to a float.
}

The \texttt{crs} attribute of the loaded \texttt{TrajDataFrame} provides the coordinate reference system, while the \texttt{parameters} attribute provides a dictionary with meta-information about the data. When we load the data from a file, \emph{scikit-mobility} adds to the \texttt{parameters} attribute the key "\texttt{from\_file}", which indicates the path of the file. 

\begin{verbatim}
Python> print(tdf.crs)
\end{verbatim}
\begin{verbatim}
{'init': 'epsg:4326'}
\end{verbatim}

\begin{verbatim}
Python> print(tdf.parameters)
\end{verbatim}
\begin{verbatim}
{'from_file': 'geolife_sample.txt.gz'}
\end{verbatim}

Once loaded, we can visualize a portion of the \texttt{TrajDataFrame} using the \texttt{print} function and the \texttt{head} function, which visualize the first five rows of the \texttt{TrajDataFrame}. Note that, since the \texttt{uid} column is present in the file, the \texttt{TrajDataFrame} created contains the corresponding column.

\begin{verbatim}
Python> print(tdf.head())
\end{verbatim}
\begin{verbatim}
         lat         lng            datetime  uid
0  39.984094  116.319236 2008-10-23 05:53:05    1
1  39.984198  116.319322 2008-10-23 05:53:06    1
2  39.984224  116.319402 2008-10-23 05:53:11    1
3  39.984211  116.319389 2008-10-23 05:53:16    1
4  39.984217  116.319422 2008-10-23 05:53:21    1
\end{verbatim}

\subsection{Flows}
Origin-destination matrices, aka flows, are another common representation of mobility data. While trajectories refer to movements of single objects, flows refer to aggregated movements of objects between a set of locations. An example of flows is the daily commuting flows between the neighbourhoods of a city. Formally, we define an origin-destination matrix as:

\begin{definition}[Origin-Destination matrix or Flows] An Origin-Destination matrix $T$ is a $n \times m$ matrix where $n$ is the number of distinct ``origin'' locations, $m$ is  the  number  of distinct  ``destination'' locations, $T_{ij}$ is  the  number  of  objects traveling from location $i$ to location $j$.
\end{definition}

In \emph{scikit-mobility}, an origin-destination matrix is described by the \texttt{FlowDataFrame} structure.
A \texttt{FlowDataFrame} is an extension of the \emph{pandas} DataFrame that has specific column names and data types. 
A row in a \texttt{FlowDataFrame} represents a flow of objects between two locations, described by three mandatory columns: \texttt{origin} (any type), \texttt{destination} (any type) and \texttt{flow} (type: integer).
Again, the user can add to a \texttt{FlowDataFrame} as many columns as they want. 

In mobility tasks, the territory is often discretized by mapping the coordinates to a spatial tessellation, i.e., a covering of the bi-dimensional space using a countable number of geometric shapes (e.g., squares, hexagons), called tiles, with no overlaps and no gaps. For instance, for the analysis or prediction of mobility flows, a spatial tessellation is used to aggregate flows of people moving among locations (the tiles of the tessellation).
For this reason, each \texttt{FlowDataFrame} is associated with a spatial tessellation, a \emph{geopandas} GeoDataFrame that contains two mandatory columns: \texttt{tile\_ID} (any type) indicates the identifier of a location; \texttt{geometry} indicates the geometric shape that describes the location on a territory (e.g., a square, an hexagon, the shape of a neighborhood).\footnote{Since a tessellation is a \emph{geopandas} \texttt{GeoDataFrame}, it supports any type of geometry (e.g., Polygon, Point). 
However, the Point geometry should be avoided because it does not correctly represent a tile of a tessellation. 
In general, Polygon and Multipolygon shapes should be preferred to describe the tiles.} 
It is important to note that each location identifier in the \texttt{origin} and \texttt{destination} columns of a \texttt{FlowDataFrame} must be present in the associated spatial tessellation. Otherwise, the library raises an exception.
Similarly, \emph{scikit-mobility} raises an exception if the type of the \texttt{origin} and \texttt{destination} columns in the \texttt{FlowDataFrame} and the type of the \texttt{tile\_ID} column in the associated tessellation are different.

The code below loads a spatial tessellation and a \texttt{FlowDataFrame} from the corresponding files. First, we import the \emph{scikit-mobility} and the \emph{geopandas} libraries.

\begin{verbatim}
Python> import skmob
Python> import geopandas as gpd
\end{verbatim}

Then, we load the \texttt{Tessellation} and the \texttt{FlowDataFrame} using the \texttt{from\_file} method of the classes GeoDataFrame and \texttt{TrajDataFrame}, respectively. Note that the \texttt{from\_file} for loading a \texttt{FlowDataFrame} requires to specify the associated \texttt{Tessellation} through the ``tessellation'' argument.

\begin{verbatim}
Python> tessellation = gpd.GeoDataFrame.from_file("NY_counties_2011.geojson")
Python> fdf = skmob.FlowDataFrame.from_file("NY_commuting_flows_2011.csv",
tessellation=tessellation, tile_id='tile_id')
\end{verbatim}

The \texttt{Tessellation} and \texttt{FlowDataFrame} have the structure shown below.

\begin{verbatim}
Python> print(tessellation.head())
\end{verbatim}
\begin{verbatim}
tile_id    population                                           geometry
0   36019       81716  POLYGON ((-74.00667 44.88602, -74.02739 44.995...
1   36101       99145  POLYGON ((-77.09975 42.27421, -77.09966 42.272...
2   36107       50872  POLYGON ((-76.25015 42.29668, -76.24914 42.302...
3   36059     1346176  POLYGON ((-73.70766 40.72783, -73.70027 40.739...
4   36011       79693  POLYGON ((-76.27907 42.78587, -76.27535 42.780...
\end{verbatim}

\begin{verbatim}
Python> print(fdf.head())
\end{verbatim}
\begin{verbatim}
     flow origin destination
0  121606  36001       36001
1       5  36001       36005
2      29  36001       36007
3      11  36001       36017
4      30  36001       36019
\end{verbatim}

\section{Trajectory preprocessing}
\label{sec:preprocessing}

As any analytical process, mobility data analysis requires data cleaning and preprocessing steps \cite{zheng2015trajectory}. The \texttt{preprocessing} module allows the user to perform three main preprocessing steps: noise filtering, stop detection, and trajectory compression. Note that, if a \texttt{TrajDataFrame} contains multiple trajectories from multiple users, the preprocessing methods automatically apply to the single trajectory and, when necessary, to the single object. 

\subsection{Noise filtering}
Trajectory data are in general noisy, usually because of recording errors like poor signal reception.  When the error associated with the coordinates of points is large, the best solution is to filter out these points. In \emph{scikit-mobility}, the method \texttt{filter} filters out a point if the speed from the previous point is higher than the parameter \texttt{max\_speed}, which is by default set to 500km/h. To use the \texttt{filter} function, we first import the preprocessing module:

\begin{verbatim}
Python> import skmob
Python> from skmob import preprocessing
\end{verbatim}

Then, we apply the filtering, setting max speed as 10 km/h, on a \texttt{TrajDataFrame} containing GPS trajectories:

\begin{verbatim}
Python> tdf = skmob.TrajDataFrame.from_file('geolife_sample.txt.gz')
Python> print('Number of points in tdf: %d\n' %len(tdf))
Python> print(tdf.head())
\end{verbatim}

\begin{verbatim}
Number of points: 217653

         lat         lng            datetime  uid
0  39.984094  116.319236 2008-10-23 05:53:05    1
1  39.984198  116.319322 2008-10-23 05:53:06    1
2  39.984224  116.319402 2008-10-23 05:53:11    1
3  39.984211  116.319389 2008-10-23 05:53:16    1
4  39.984217  116.319422 2008-10-23 05:53:21    1
\end{verbatim}

\begin{verbatim}
Python> ftdf = preprocessing.filtering.filter(tdf, max_speed_kmh=10.)
Python> print("Number of points in ftdf: %d" %len(ftdf))
Python> print("Number of filtered points: %d\n" %(len(tdf) - len(ftdf)))
Python> print(ftdf.head())
\end{verbatim}
\begin{verbatim}
Number of points in ftdf: 108779
Number of filtered points: 108874

         lat         lng             datetime  uid
0  39.984094  116.319236  2008-10-23 05:53:05  1
1  39.984211  116.319389  2008-10-23 05:53:16  1
2  39.984217  116.319422  2008-10-23 05:53:21  1
3  39.984555  116.319728  2008-10-23 05:53:43  1
4  39.984579  116.319769  2008-10-23 05:53:48  1
\end{verbatim}

As we can see, 108,874 points out of 217,653 are filtered out. 
The intensity of the filter is controlled by the \texttt{max\_speed} parameter. 
The lower the value, the more intense the filter is.

\subsection{Stop detection}
Some points in a trajectory can represent Point-Of-Interests (POIs) such as schools, restaurants, and bars, or they can represent user-specific places such as home and work locations. 
These points are usually called \emph{Stay Points} or \emph{Stops}, and they can be detected in different ways. 
A common approach is to apply spatial clustering algorithms to cluster trajectory points by looking at their spatial proximity \cite{hariharan2004project}. 
In \emph{scikit-mobility}, the \texttt{stops} function, contained in the \texttt{detection} module, finds the stay points visited by an object. For instance, to identify the stops where the object spent at least \texttt{minutes\_for\_a\_stop} minutes within a distance \texttt{spatial\_radius\_km} $\times$ \texttt{stop\_radius\_factor}, from a given point, we can use the following code:

\begin{verbatim}
Python> from preprocessing import detection
Python> stdf = detection.stops(ctdf, stop_radius_factor=0.5, minutes_for_a_stop=20.0, 
        spatial_radius_km=0.2)
\end{verbatim}
\begin{verbatim}
         lat         lng            datetime  uid    leaving_datetime
0  39.978253  116.327275 2008-10-23 06:01:05    1 2008-10-23 10:32:53
1  40.013819  116.306532 2008-10-23 11:10:09    1 2008-10-23 23:46:02
2  39.978950  116.326439 2008-10-24 00:12:30    1 2008-10-24 01:48:57
3  39.981316  116.310181 2008-10-24 01:56:47    1 2008-10-24 02:28:19
4  39.981451  116.309505 2008-10-24 02:28:19    1 2008-10-24 03:18:23
\end{verbatim}

As shown in the code snippet, a new column \texttt{leaving\_datetime} is added to the \texttt{TrajDataFrame} in order to indicate the time when the user left the stop location.

\subsection{Trajectory compression}
The goal of trajectory compression is to reduce the number of trajectory points while preserving the structure of the trajectory. This step is generally applied right after the stop detection step, and it results in a significant reduction of the number of trajectory points. In \emph{scikit-mobility}, we can use one of the methods in the \texttt{compression} module under the preprocessing module. For instance, to merge all the points that are closer than $0.2 km$ from each other, we can use the following code:

\begin{verbatim}
Python> from preprocessing import compression
Python> print(ftdf.head())
\end{verbatim}
\begin{verbatim}
         lat         lng            datetime  uid
0  39.984094  116.319236  2008-10-23 05:53:05  1
1  39.984211  116.319389  2008-10-23 05:53:16  1
2  39.984217  116.319422  2008-10-23 05:53:21  1
3  39.984555  116.319728  2008-10-23 05:53:43  1
4  39.984579  116.319769  2008-10-23 05:53:48  1
\end{verbatim}

\begin{verbatim}
Python> ctdf = compression.compress(ftdf, spatial_radius_km=0.2)
\end{verbatim}
\begin{verbatim}
   lat        lng         datetime            uid
0  39.984334  116.320778  2008-10-23 05:53:05  1
1  39.979642  116.322241  2008-10-23 05:58:33  1
2  39.978051  116.327538  2008-10-23 06:01:47  1
3  39.970511  116.341455  2008-10-23 10:32:53  1
\end{verbatim}

Once compressed, the trajectory will present a smaller number of points, allowing then an easy plotting of them by using the data visualization functionalities of \emph{scikit-mobility} described in Section \ref{sec:plotting}.
Table \ref{tab:preproc} lists the available methods for trajectory preprocessing.

\begin{table}[]
    \centering
    \begin{tabular}{c|p{10cm}}
    \bf method & \bf description \\
    \hline
    \texttt{clustering.cluster} & Cluster the stops of each individual in a \texttt{TrajDataFrame}. 
    Uses DBSCAN \cite{hariharan2004project}\\
    \hline
    \texttt{compression.compress} & Reduce the number of points of each individual in a \texttt{TrajDataFrame} with median coordinates within a radius \cite{zheng2015trajectory}\\
    \hline
    \texttt{detection.stops} & Detect the stops for each individual in a \texttt{TrajDataFrame} with a time threshold \cite{hariharan2004project, zheng2015trajectory}\\
    \hline
    \texttt{filtering.filter} & For each trajectory, filters out the noise or outlier points \cite{zheng2015trajectory}
    \end{tabular}
    \caption{Trajectory preprocessing methods implemented in \emph{scikit-mobility}.}
    \label{tab:preproc}
\end{table}

%\subsection{Trajectory segmentation}
%Many datasets, such as Location-Based Social Networks (LBSNs), contain trajectories that are a sequences of locations where the user decided to voluntary register his/her presence. Thus, the trajectories need to be divided into segment for further process. This process not only reduce the computational complexity for each single trajectory, but it also allows for a deeper learning of the data, e.g., subtrajectory pattern mining.
%As described in \cite{zheng2015trajectory}, there are three categories of segmentation method that can be found in the library: (i) time interval, which divides the trajectory when the difference between points is larger than a given time threshold.  (ii) shape of the trajectory, where the sequence of locations is partitioned by the turning points with heading direction changing over a threshold. (iii) semantic meanings, which segment the trajectory based on the stay points it contains. Depending on the application, only some of the stay points can be kept. For instance, if we want to predict the next location in a taxi ride, we construct the trajectory only considering the pick-up and drop-off points.

\section{Plotting}
\label{sec:plotting}

% \begin{CodeChunk}
% \begin{CodeInput}
% Python> 
% \end{CodeInput}
% \begin{CodeOutput}

% \end{CodeOutput}
% \end{CodeChunk}

One of the use cases for \emph{scikit-mobility} is the exploratory data analysis of mobility data sets, which includes the visualization of trajectories and flows. 
To this end, both \texttt{TrajDataFrame} and \texttt{FlowDataFrame} have methods that allow the user to produce interactive visualizations generated using the library \ee{folium} \cite{folium}.
The choice of \ee{folium} is motivated by the fact that, given the complexity of mobility data, the user may need to zoom in/out and interact with the components of trajectories, flows and tessellations. 
This type of interaction would be not possible with static plotting libraries, such as \ee{matplotlib}. 
The user can save an interactive plot in a \texttt{.html} file or they can take a screenshot to save it on a \texttt{.png} file.

\subsection{Visualizing trajectories}\label{sec:plot_traj}

A \texttt{TrajDataFrame} has three main plotting methods: \texttt{plot\_trajectory} plots a line connecting the trajectory points on a map; \texttt{plot\_stops} plots the location of stops on a map; and 
\texttt{plot\_diary} plots the sequence of visited locations over time. 

\subsubsection{Plot trajectories}

The \texttt{TrajDataFrame}'s method \texttt{plot\_trajectory} plots the time-ordered trajectory points connected by straight lines on a map. 
If the column \texttt{uid} is present and contains more than one object, the trajectory points are first grouped by \texttt{uid} and then sorted by \texttt{datetime}. 
Large \texttt{TrajDataFrame}s with many points can be computationally intensive to visualize. Two arguments can be used to reduce the amount of data to plot:  \texttt{max\_users} (type: \texttt{int}, default: 10) limits the number of objects whose trajectories should be plotted, while 
\texttt{max\_points} (type: \texttt{int}, default: 1000) limits the number of trajectory points per object to plot, i.e., if necessary, an object's trajectory will be down-sampled and at most \texttt{max\_points} points will be plotted. 
The plot style can be customized via arguments to specify the color, weight, and opacity of the trajectory lines, as well as the type of map tiles to use. 
The user can also plot markers denoting the start points and the end points of the trajectory.

The \texttt{plot\_trajectory} method, as well as all the other plotting methods, return a \texttt{folium.Map} object, which can be used by other \ee{folium} and \emph{scikit-mobility} functions in order to visualize additional data on the same map. A \texttt{folium.Map} object can be passed to a plotting method via the argument \texttt{map\_f} (default: \texttt{None}, which means that the mobility data are plotted on a new map).

An example of plot generated by the \texttt{plot\_trajectory} method is shown below: 

\begin{verbatim}
Python> import skmob
Python> tdf = skmob.TrajDataFrame.from_file('geolife_sample.txt.gz')
Python> map_f = tdf.plot_trajectory(max_users=1, hex_color='#000000')
Python> map_f
\end{verbatim}
\includegraphics[width=0.75\columnwidth]{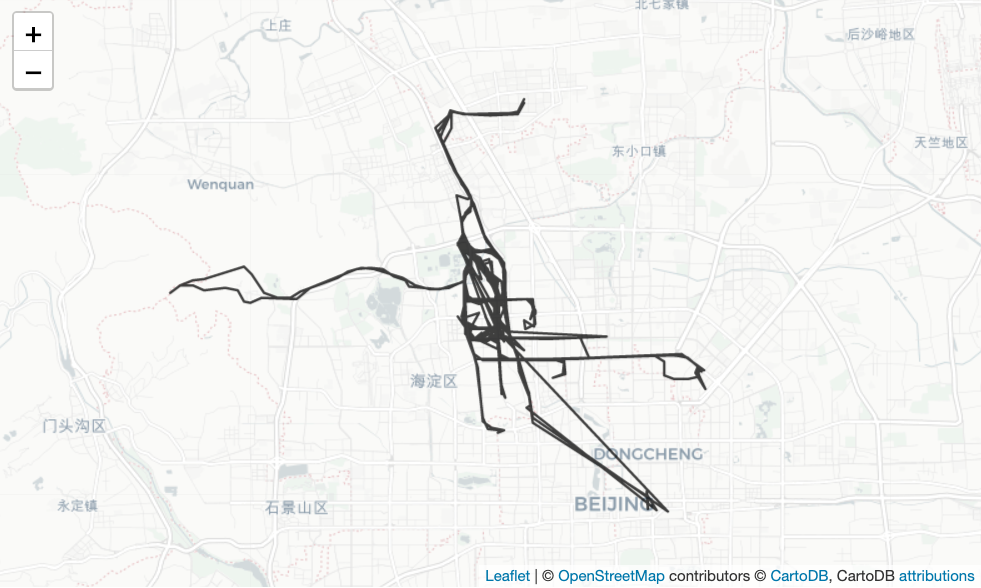}

Note that if trajectories represent abstract mobility, such as movements extracted from social media posts or mobile phone calls, straight lines may appear that do not take into account walls, buildings and similar structures on the road network.

By default, a \texttt{TrajDataFrame} represents the full mobility of a set of individuals, i.e., covering the entire period of observation (e.g., one month). The user can split the trajectory of an individual using preprocessing functions, such as the \texttt{detection.stops} function (Section \ref{sec:preprocessing}), and then split the whole trajectory into sub trajectories, adding a proper column to identify them (i.e., the \texttt{tid} column). 
At this point, the user may visualize the portion of the \texttt{TrajDataFrame} selecting for values of the created column.

\subsubsection{Plot stops}

The \texttt{TrajDataFrame}'s method \texttt{plot\_stops} plots the locations of the stops as markers on a map. 
This method requires a \texttt{TrajDataFrame} with the column \texttt{constants.LEAVING\_DATETIME}, which is created by the \emph{scikit-mobility} functions to detect stops (see \ref{sec:preprocessing}). 
The argument \texttt{max\_users} (type: \texttt{int}, default: 10) limits the number of objects whose stops should be plotted. 
The plot style can be customized via arguments to specify the color, radius, and opacity of the markers, as well as the type of the map tiles to use. 
The argument \texttt{popup} (default: \texttt{False}) allows enhancing the plot's interactivity displaying popup windows that appear when the user clicks on a marker. A stop's popup window includes information like coordinates, object's \texttt{uid}, arrival, and leaving times. 

The method returns a \texttt{folium.Map} object, which can be used by other \ee{folium} and \emph{scikit-mobility} functions in order to visualize additional data on the same map. A \texttt{folium.Map} object can be passed to \texttt{plot\_stops} via the argument \texttt{map\_f} (default: \texttt{None}, which means that the stops are plotted on a new map). 

We show below an example of a plot generated by the \texttt{plot\_stops} method. Note that if the \texttt{cluster} column is present in the \texttt{TrajDataFrame}, as it happens for instance when the \texttt{cluster} method is applied (Section \ref{sec:preprocessing}), the stops are automatically colored according to the value of that column (so as to identify different clusters of stops).

\begin{verbatim}
Python> from skmob.preprocessing import detection, clustering 
Python> tdf = skmob.TrajDataFrame.from_file('geolife_sample.txt.gz')
Python> stdf = detection.stops(tdf)
Python> cstdf = clustering.cluster(stdf)
Python> cstdf.plot_stops(max_users=1, map_f=mapf)
\end{verbatim}
\includegraphics[width=0.75\columnwidth]{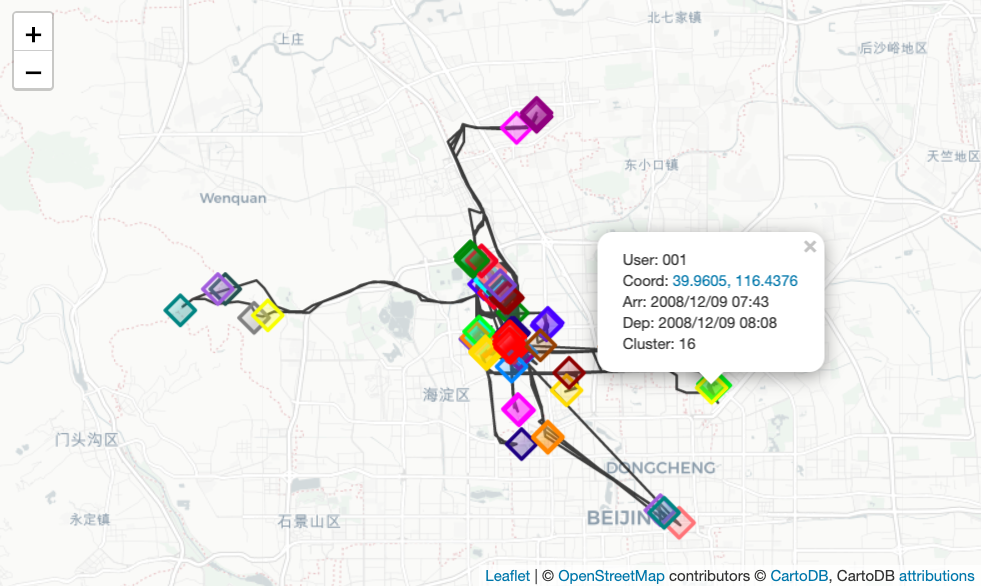}

\subsubsection{Plot diary}

The \texttt{TrajDataFrame}'s method \texttt{plot\_diary} plots the time series of the locations visited by an object. 
If the column \texttt{uid} is present, one object ID must be specified via the argument \texttt{user}. 
This method requires a \texttt{TrajDataFrame} with the column \texttt{constants.CLUSTER}, which is created by the \emph{scikit-mobility} functions to cluster stops (see \ref{sec:preprocessing}). 

The plot displays time on the \ee{x} axis and shows a series of rectangles of different colors that represent the object's visits to the various stops. 
The length of a rectangle denotes the duration of the visit: the left edge marks the arrival time, the right edge marks the leaving time. 
The color of a rectangle denotes the stop's cluster: visits to stops that belong to the same cluster have the same color (the color code is consistent with the one used by the method \texttt{plot\_stops}). 
A white rectangle indicates that the object is moving. 

We show below an example of a plot generated by the \texttt{plot\_diary} method:

\begin{verbatim}
Python> cstdf.plot_diary(user='001')
\end{verbatim}
\includegraphics[width=0.9\columnwidth]{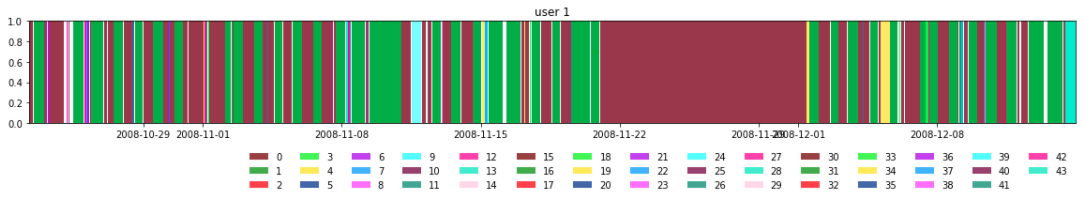}

The user can compare multiple moving objects plotting their diaries next to each other:

\begin{verbatim}
Python> ax = cstdf.plot_diary(1)
Python> ax = cstdf.plot_diary(5, legend=True)
\end{verbatim}
\includegraphics[width=0.9\columnwidth]{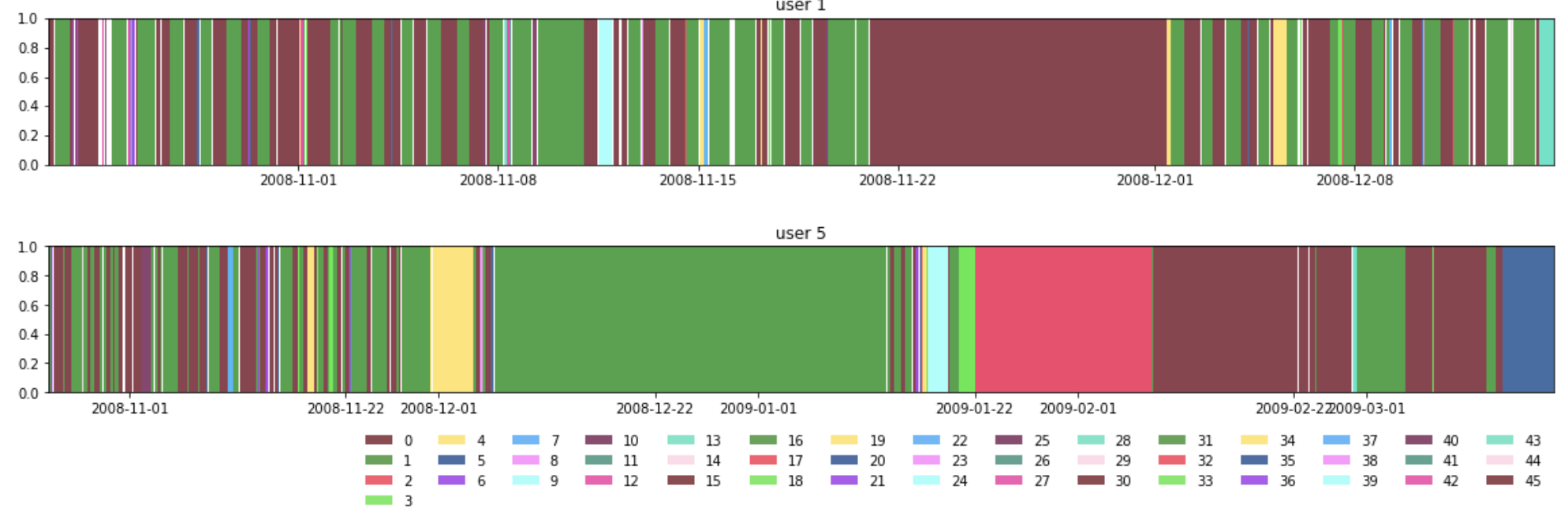}

\subsection{Visualizing flows}\label{sec:plot_flow}

A \texttt{FlowDataFrame} has two main plotting methods: 
\texttt{plot\_tessellation} plots the tessellation's tiles on a geographic map 
and 
\texttt{plot\_flows} plots, on a geographic map, the lines connecting the centroids of the tessellation's tiles between which flows are present. 

\subsubsection{Plot tessellation}

The \texttt{FlowDataFrame}'s method \texttt{plot\_tessellation} plots the \texttt{GeoDataFrame} associated with a \texttt{FlowDataFrame} on a geographic map. 
Large tessellations with many tiles can be computationally intensive to visualize. The argument \texttt{maxitems} can be used to limit the number of tiles to plot (default: -1, which means that all tiles are displayed). 

The plot style can be customized via arguments to specify the color and opacity of the tiles, as well as the type of map tiles to use. 
The argument \texttt{popup\_features} (type: \texttt{list}, default: \texttt{[constants.TILE\_ID]}) allows to enhance the plot's interactivity displaying popup windows that appear when the user clicks on a tile and includes information contained in the columns of the tessellation's \texttt{GeoDataFrame} specified in the argument's list.

The method returns a \texttt{folium.Map} object, which can be used by other \ee{folium} and \emph{scikit-mobility} functions in order to visualize additional data on the same map. A \texttt{folium.Map} object can be passed to \texttt{plot\_flows} via the argument \texttt{map\_osm} (default: \texttt{None}, which means that the tessellation is plotted on a new map). 

We show below an example of a plot generated by the \texttt{plot\_tessellation} method:

\begin{verbatim}
Python> import geopandas as gpd
Python> from skmob import FlowDataFrame
Python> tessellation = gpd.GeoDataFrame.from_file('./NY_counties_2011.geojson')
Python> fdf = FlowDataFrame.from_file('./NY_commuting_flows_2011.csv', 
                                       tessellation=tessellation)
Python> fdf.plot_tessellation(popup_features=['tile_ID', 'population'])
\end{verbatim}
\includegraphics[width=0.75\columnwidth]{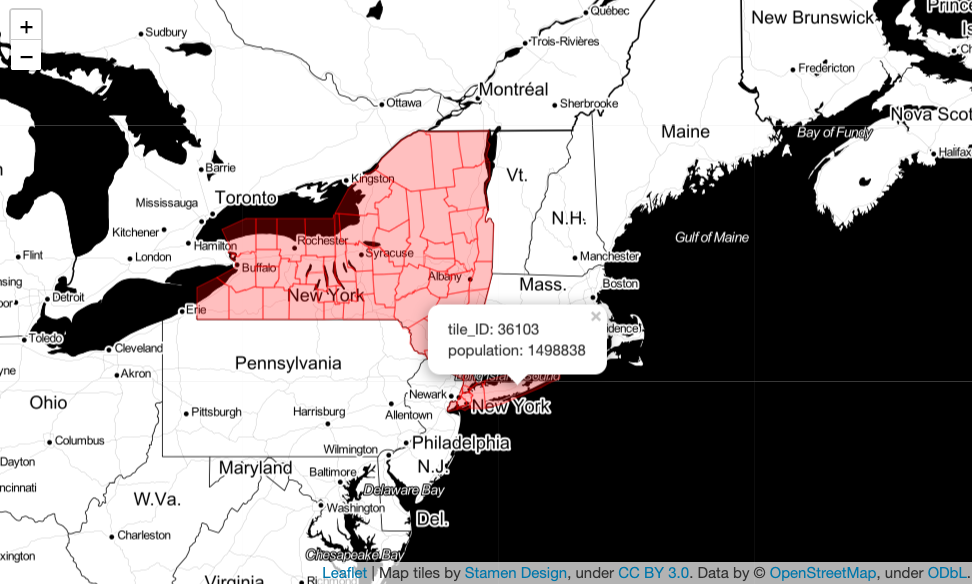}

\subsubsection{Plot flows}

The \texttt{FlowDataFrame}'s method \texttt{plot\_flows} plots the flows on a geographic map as lines between the centroids of the tiles in the \texttt{FlowDataFrame}'s tessellation. 
Large \texttt{FlowDataFrame}{}s with many origin-destination pairs can be computationally intensive to visualize. 
The argument \texttt{min\_flow} (type: integer, default: 0) can be used to specify that only flows larger than \texttt{min\_flow} should be displayed. 
The thickness of each line is a function of the flow and can be specified via the arguments \texttt{flow\_weight}, \texttt{flow\_exp} and \texttt{style\_function}. 
The plot style can be further customized via arguments to specify the color and opacity of the flow lines, as well as the type of map tiles to use. 
The arguments \texttt{flow\_popup} and \texttt{tile\_popup} allow to enhance the plot's interactivity displaying popup windows that appear when the user clicks on a flow line or a circle in an origin location, respectively, and include information on the flow or the flows from a location. 
The method returns a \texttt{folium.Map} object, which can be used by other \ee{folium} and \emph{scikit-mobility} functions in order to visualize additional data on the same map. A \texttt{folium.Map} object can be passed to \texttt{plot\_flows} via the argument \texttt{map\_f} (default: \texttt{None}, which means that the flows are plotted on a new map). 

We show below an example of a plot generated by the \texttt{plot\_flows} method:

\begin{verbatim}
Python> fdf.plot_flows(min_flow=50)
\end{verbatim}
\includegraphics[width=0.75\columnwidth]{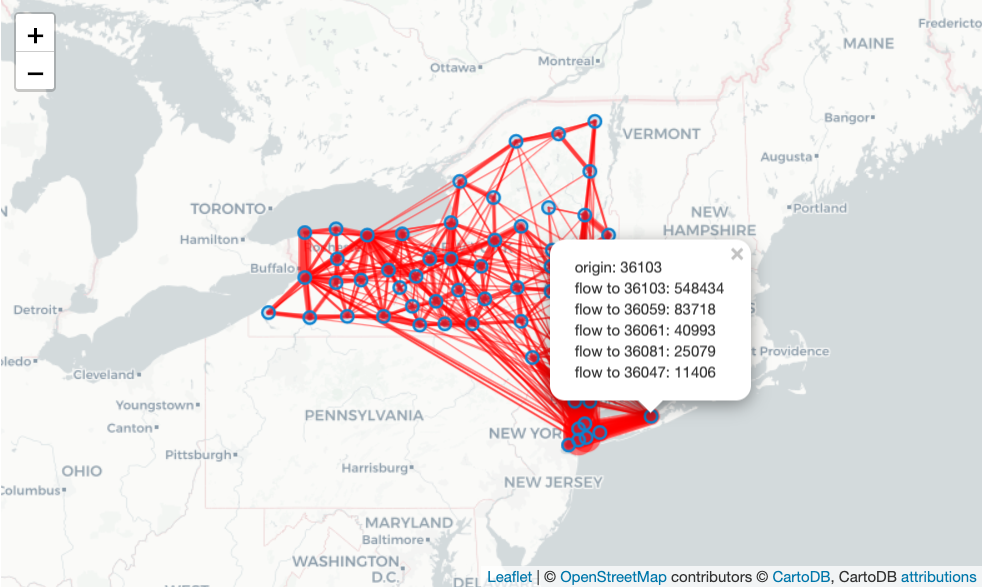}

Table \ref{tab:plot} lists the plotting functions available in the library.

The user can also visualize the tessellation and the flows in the same plot:

\begin{verbatim}
Python> map_f = fdf.plot_tessellation(popup_features=['tile_id','population'], 
                    style_func_args={'color': 'red'})
Python> fdf.plot_flows(map_f=map_f, min_flow=50)
\end{verbatim}
\includegraphics[width=0.75\columnwidth]{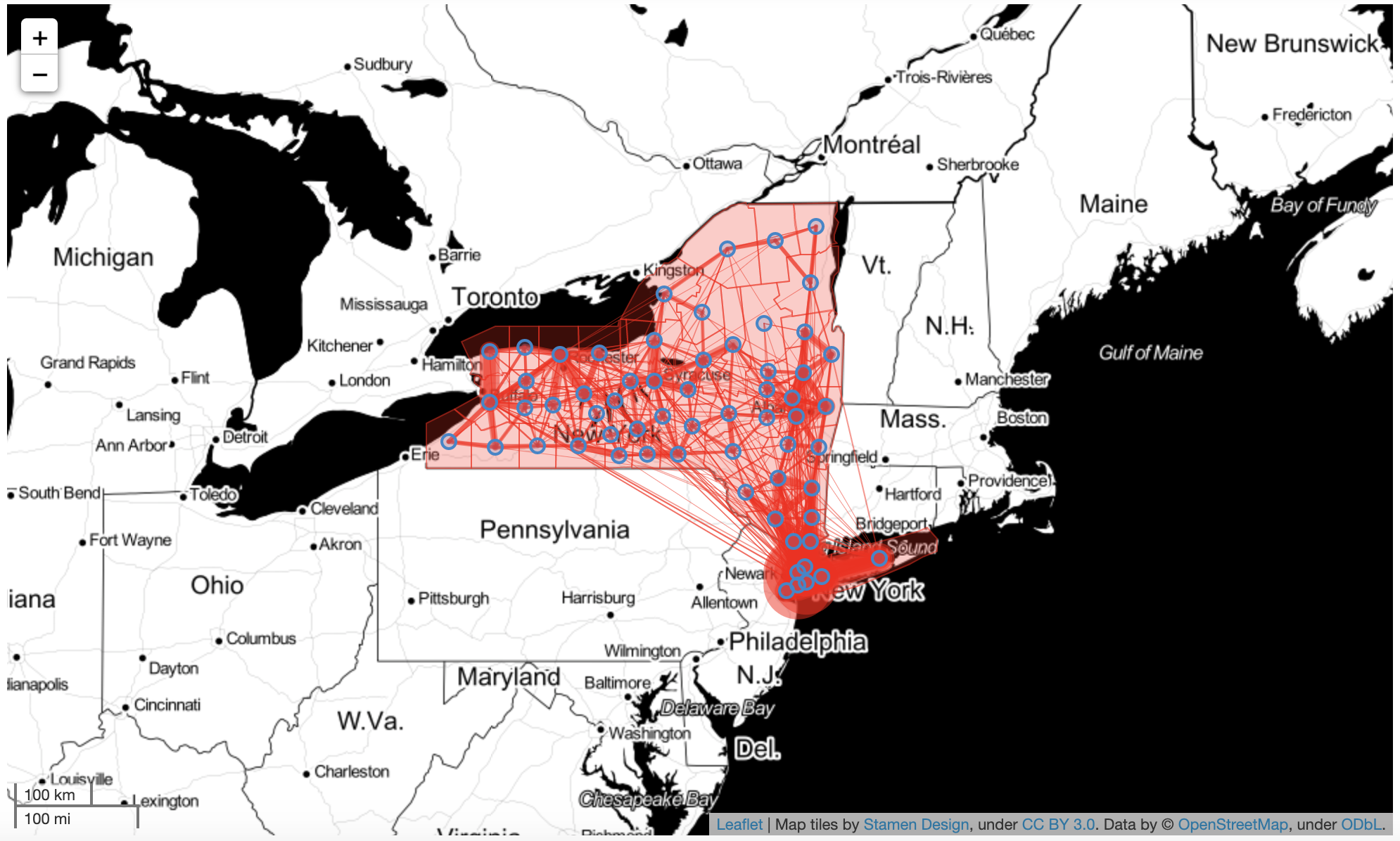}

\begin{table}[]
    \centering
    \begin{tabular}{c|p{12cm}}
    \bf method & \bf description \\
    \hline
    \texttt{plot\_diary} & plot a mobility diary of an individual \cite{hariharan2004project}\\
    \hline
    \texttt{plot\_stops} & plot the stops in the \texttt{TrajDataFrame} on a \emph{folium} map \\
    \hline
    \texttt{plot\_trajectory} & plot the trajectories on a \emph{folium} map\\
    \hline
    \texttt{plot\_flows} & plot the flows of a \texttt{FlowDataFrame} on a \emph{folium} map\\
    \hline
    \texttt{plot\_tessellation} & plot the spatial tessellation on a \emph{folium} map\\
    \end{tabular}
    \caption{Plotting methods implemented in \emph{scikit-mobility}.}
    \label{tab:plot}
\end{table}

\section{Mobility measures}
\label{sec:measures}
In the last decade, several measures have been proposed to capture the patterns of human mobility, both at the individual and collective levels. Individual measures summarize the mobility patterns of a single moving object, while collective measures summarize mobility patterns of a population as a whole.
For instance, the so-called radius of gyration \cite{gonzalez08understanding} and its variants \cite{pappalardo2015returners} quantify the characteristic distance traveled by an individual, while several measures inspired by the Shannon entropy have been proposed to quantify the predictability of an individual's movements \cite{song2010limits}. 

\emph{scikit-mobility} provides a wide set of mobility measures, each implemented as a function that takes in input a \texttt{TrajDataFrame} and outputs a \emph{pandas} DataFrame. Individual and collective measures are implemented the in \texttt{skmob.measure.individual} and the \texttt{skmob.measures.collective} modules, respectively. 

The code below computes two measures: the distances traveled by the objects and their radius of gyration. First, we import the two functions from the library. 

\begin{verbatim}
Python> import skmob
Python> from skmob.measures.individual import jump_lengths, radius_of_gyration
\end{verbatim}

Then, we invoke the two functions on the \texttt{TrajDataFrame}, respectively. 

\begin{verbatim}
Python> jl_df = jump_lengths(tdf)
Python> rg_df = radius_of_gyration(tdf)
\end{verbatim}

The output of the functions is a \emph{pandas} DataFrame with two columns: \texttt{uid} contains the identifier of the object; the second column, the name of which corresponds to the name of the invoked function, contains the computed measure for that object. 
For example, in the DataFrame \texttt{jl\_df}, the column \texttt{jump\_length} of the DataFrame contains a list of all distances traveled by that object. 

\begin{verbatim}
Python> print(jl_df.head())
\end{verbatim}
\begin{verbatim}
   uid                                       jump_lengths
0    0  [19.640467328877936, 0.0, 0.0, 1.7434311010381...
1    1  [6.505330424378251, 46.75436600375988, 53.9284...
2    2  [0.0, 0.0, 0.0, 0.0, 3.6410097195943507, 0.0, ...
3    3  [3861.2706300798827, 4.061631313492122, 5.9163...
4    4  [15511.92758595804, 0.0, 15511.92758595804, 1....
\end{verbatim}

Similary, in the DataFrame \texttt{rg\_df} the column \texttt{radius\_of\_gyration} contains the radius of gyration for that object.

\begin{verbatim}
Python> print(rg_df.head())
\end{verbatim}
\begin{verbatim}
   uid  radius_of_gyration
0    0         1564.436792
1    1         2467.773523
2    2         1439.649774
3    3         1752.604191
4    4         5380.503250
\end{verbatim}

Note that, if the optional column \texttt{uid} is not present in the input \texttt{TrajDataFrame}, a simple Python structure is outputted instead of the \emph{pandas} DataFrame (e.g., a list for function \texttt{jump\_lengths} and a float for function \texttt{radius\_of\_gyration}).

Collective measures are used in a similar way. The code below computes a collective measure - the number of visits per location (by any object). First, we import the function.

\begin{verbatim}
Python> import skmob
Python> from skmob.measures.collective import visits_per_location
\end{verbatim}

Then, we invoke the function on the \texttt{TrajDataFrame}.

\begin{verbatim}
Python> vpl_df = visits_per_location(tdf)
\end{verbatim}

As for the individual measures, the output of the functions is a \emph{pandas} DataFrame. The format of this DataFrame depends on the measures. For example, in the DataFrame \texttt{vpl\_df} there are three columns: \texttt{lat} and \texttt{lng} indicate the coordinates of a location, and \texttt{n\_visits} indicate the number of visits to that location in the \texttt{TrajDataFrame}.

\begin{verbatim}
Python> print(vpl_df.head())
\end{verbatim}
\begin{verbatim}
         lat         lng  n_visits
0  39.739154 -104.984703      3392
1  37.580304 -122.343679      2248
2  39.099275  -76.848306      1715
3  39.762146 -104.982480      1442
4  40.014986 -105.270546      1310
\end{verbatim}

Table \ref{tab:individual_measures} and Table \ref{tab:collective_measures} list the available individual and collective measures, respectively.

\begin{table}[]
    \centering
    \begin{tabular}{c|p{10cm}}
    \bf individual measure &  \bf description \\
    \hline
    \texttt{radius\_of\_gyration} & characteristic distance travelled by an individual \cite{gonzalez08understanding}\\
    \hline
    \texttt{k\_radius\_of\_gyration} & characteristic distance travelled by an individual between their $k$ most frequent locations \cite{pappalardo2015returners}\\
    \hline
    \texttt{random\_entropy} & degree of predictability of an individual's whereabouts if each location is visited with equal probability \cite{song2010limits}. \\
    \hline
    \texttt{uncorrelated\_entropy} & historical probability that a location was visited by an individual \cite{song2010limits} \\
    \hline
    \texttt{real\_entropy} & mobility entropy of an individual considering also the order in which locations were visited  \cite{song2010limits} \\
    \hline
    \texttt{jump\_length} & distances traveled by an individual \cite{brockmann2006scaling}\\
    \hline
    \texttt{maximum\_distance} & maximum distance traveled by an individual \cite{williams2014measures} \\
    \hline
    \texttt{distance\_straight\_line} & sum of the distances traveled by an individual \cite{williams2014measures}\\
    \hline
    \texttt{waiting\_times}& inter-times between the movements of an individual \cite{song2010modelling} \\
    \hline
    \texttt{number\_of\_locations} & number of distinct locations visited by an individual \\
    \hline
    \texttt{home\_location} & location most visited by an individual during nighttime \cite{Phithakkitnukoon2012socio} \\
    \hline
    \texttt{max\_distance\_from\_home} & maximum distance from home traveled by an individual \cite{canzian2015trajectories} \\
    \hline
    \texttt{number\_of\_visits} & number of visits to any location by an individual \\
    \hline
    \texttt{location\_frequency} & visitation frequency of each location of an individual \cite{song2010modelling}\\
    \hline
    \texttt{individual\_mobility\_network} & individual mobility network of an individual \cite{bagrow2012mesoscopic, rinzivillo2014purpose}\\
    \hline
    \texttt{recency\_rank} & recency rank of the locations of an individual \cite{barbosa2015effect}\\
    \hline
    \texttt{frequency\_rank} & frequency rank of the locations of an individual \cite{barbosa2015effect}\\
    \end{tabular}
    \caption{Individual measures implemented in \emph{scikit-mobility}.}
    \label{tab:individual_measures}
\end{table}

\begin{table}[]
    \centering
    \begin{tabular}{c|p{10cm}}
    \bf collective measure &  \bf description \\
    \hline
    \texttt{random\_location\_entropy} & the random entropy of locations with respect to individual visits\\
    \texttt{uncorrelated\_location\_entropy} & the historical probability that an individual visited location \cite{cho2011friendship}\\
    \hline
    \texttt{mean\_square\_displacement} & the mean square displacement traveled by the individuals after a time \cite{brockmann2006scaling}\\
    \hline
    \texttt{visits\_per\_location} & number of visits per location \cite{pappalardo2018data}\\
    \hline
    \texttt{homes\_per\_location} & number of homes per location \cite{pappalardo2018data} \\
    \hline
    \texttt{visits\_per\_time\_unit} & number of visits to any location per time unit \cite{pappalardo2018data} \\
    \hline
    \texttt{origin\_destination\_matrix} & origin-destination matrix from the trajectories of the individuals \cite{calabrede2011ODM}
    \end{tabular}
    \caption{Collective measures implemented in \emph{scikit-mobility}.}
    \label{tab:collective_measures}
\end{table}

\section{Individual Generative Algorithms}
\label{sec:individual_models}
The goal of generative algorithms of human mobility is to create a population of agents whose mobility patterns are statistically indistinguishable from those of real individuals \cite{pappalardo2018data}. A generative algorithm typically generates a synthetic trajectory corresponding to a single moving object, assuming that an object is independent of the others. 
\emph{scikit-mobility} implements the most common individual generative algorithms, such as the Exploration and Preferential Return model \cite{song2010modelling} and its variants \cite{pappalardo2016human, barbosa2015effect, alessandretti2018evidence}, and DITRAS \cite{pappalardo2018data}.
Each generative algorithm is a python class. First, we instantiate the algorithm. Then we invoke the \texttt{generate} method to start the generation of synthetic trajectories.

The code below shows the code to generate a \texttt{TrajDataFrame} describing the synthetic trajectory of 1000 agents that move between the locations of a \texttt{Tessellation} and for a period specified in the input. First, we import the class of the generative algorithm (\texttt{DensityEPR}) from the library. 

\begin{verbatim}
Python> import skmob
Python> import pandas as pd
Python> import geopandas as gpd
Python> from skmob.models.epr import DensityEPR
\end{verbatim}

Then, we load the spatial tessellation on which the agents have to move from a file as a \texttt{Tessellation} object, and we specify the start and end times of the simulation as pandas datetime objects.

\begin{verbatim}
Python> tessellation = gpd.GeoDataFrame.from_file("NY_counties_2011.geojson")
Python> start_time = pd.to_datetime('2019/01/01 08:00:00')
Python> end_time = pd.to_datetime('2019/01/14 08:00:00')
\end{verbatim}

Finally, we instantiate the \texttt{DensityEPR} model and start the simulation through the \texttt{generate} method, which takes in input the start and end times, the \texttt{Tessellation}, the number of agents, and other model-specific parameters. The output of the simulation is a \texttt{TrajDataFrame} containing the trajectory of the 1000 agents.

\begin{verbatim}
Python> depr = DensityEPR()
Python> tdf = depr.generate(start_time, end_time, tessellation, n_agents=1000, 
                 relevance_column='population', random_state=42)
Python> print(tdf.head())
\end{verbatim}
\begin{verbatim}
   uid                   datetime        lat        lng
0    1 2019-01-01 08:00:00.000000  42.393730 -76.875204
1    1 2019-01-01 08:36:25.019263  42.452018 -76.473618
2    1 2019-01-01 09:10:52.828149  42.393730 -76.875204
3    1 2019-01-02 03:14:59.370208  42.702464 -78.224637
4    1 2019-01-02 03:40:17.509278  44.592993 -74.303615
\end{verbatim}

\section{Collective Generative Algorithms}
\label{sec:collective_models}

Collective generative algorithms estimate spatial flows between a set of discrete locations. 
Examples of spatial flows estimated with collective generative algorithms include commuting trips between neighborhoods, migration flows between municipalities, freight shipments between states, and phone calls between regions \cite{barbosa2018human}. 

In \emph{scikit-mobility}, a collective generative algorithm takes in input a \texttt{Tessellation}.

To be a valid input for a collective algorithm, the \texttt{Tessellation} should contain two columns, \texttt{geometry} and \texttt{relevance}, which are necessary to compute the two variables used by collective algorithms: the distance between tiles and the importance (aka ``attractiveness'') of each tile. 
A collective algorithm produces a \texttt{FlowDataFrame} that contains the generated flows and the \texttt{Tessellation} of which is the one specified as the algorithm's input. 

\emph{scikit-mobility} implements the most common collective generative algorithms: the Gravity model~\cite{zipf1946p,wilson1971family} and the Radiation model~\cite{simini2012universal}. 
We illustrate how to work with generative algorithms in \emph{scikit-mobility} with an example based on the Gravity model. 

The class \texttt{Gravity}, implementing the Gravity model, has two main methods: \texttt{fit}, which calibrates the model's parameters using a training \texttt{FlowDataFrame}; and \texttt{generate}, which generates the flows on a given tessellation. 
The following code shows how to use both methods to estimate the commuting flows between the counties in the state of New York. First, we load the tessellation from a file: 

\begin{verbatim}
Python> import skmob
Python> import geopandas as gpd
Python> tessellation = gpd.GeoDataFrame.from_file("NY_counties_2011.geojson")
Python> print(tessellation.head())
\end{verbatim}
\begin{verbatim}
  tile_id  population                                           geometry
0   36019       81716  POLYGON ((-74.006668 44.886017, -74.027389 44....
1   36101       99145  POLYGON ((-77.099754 42.274215, -77.0996569999...
2   36107       50872  POLYGON ((-76.25014899999999 42.296676, -76.24...
3   36059     1346176  POLYGON ((-73.707662 40.727831, -73.700272 40....
4   36011       79693  POLYGON ((-76.279067 42.785866, -76.2753479999...
\end{verbatim}

The tessellation contains the column \texttt{population},  used as relevance variable for each tile (county). 
Next, we load the observed commuting flows between the counties from file: 

\begin{verbatim}
Python> import skmob
Python> fdf = skmob.FlowDataFrame.from_file("NY_commuting_flows_2011.csv", 
        tessellation=tessellation)
Python> print(fdf.head())
\end{verbatim}
\begin{verbatim}
     flow origin destination
0  121606  36001       36001
1       5  36001       36005
2      29  36001       36007
3      11  36001       36017
4      30  36001       36019
\end{verbatim}

Let us use the observed flows to fit the parameters of a singly-constrained gravity model with the power-law deterrence function (for more details on the gravity models see~\cite{barbosa2018human}. 
First, we instantiate the model: 

\begin{verbatim}
Python> from skmob.models.gravity import Gravity
Python> gravity = Gravity(gravity_type='singly constrained')
Python> print(gravity)
\end{verbatim}
\begin{verbatim}
Gravity(name="Gravity model", deterrence_func_type="power_law", 
deterrence_func_args=[-2.0], origin_exp=1.0, destination_exp=1.0,
gravity_type="singly constrained")
\end{verbatim}

Then we call the method \texttt{fit} to fit the parameters from the previously loaded \texttt{FlowDataFrame}: 

\begin{verbatim}
Python> gravity.fit(fdf, relevance_column='population')
Python> print(gravity)
\end{verbatim}
\begin{verbatim}
Gravity(name="Gravity model", deterrence_func_type="power_law",
deterrence_func_args=[-1.99471520], origin_exp=1.0, destination_exp=0.64717595,
gravity_type="singly constrained")
\end{verbatim}

Finally, we use the fitted model to generate the flows on the same tessellation. Setting the argument \texttt{out\_format="probabilities"} we specify that in the column \texttt{flow} of the returned \texttt{FlowDataFrame} we want the probability to observe a unit flow (trip) between two tiles. 

\begin{verbatim}
Python> fdf_fitted = gravity.generate(tessellation, 
        relevance_column='population', out_format='probabilities')
Python> print(fdf_fitted.head())
\end{verbatim}
\begin{verbatim}
  origin destination      flow
0  36019       36101  0.004387
1  36019       36107  0.003702
2  36019       36059  0.019679
3  36019       36011  0.006894
4  36019       36123  0.002292
\end{verbatim}

Table \ref{tab:genmodels} lists the generative models available in the library.

\begin{table}[]
    \centering
    \begin{tabular}{c|p{10cm}}
        \bf generative model & \bf description \\
        \hline
        \texttt{DensityEPR} & Density Exploration and Preferential Return model \cite{pappalardo2015returners,pappalardo2016human}\\
        \hline
        \texttt{SpatialEPR} & Spatial Exploration and Preferential Return model \cite{pappalardo2015returners,pappalardo2016human,song2010modelling}\\
        \hline
        \texttt{Ditras} & DIary-based TRAjectory Simulator modelling framework \cite{pappalardo2018data} \\
        \hline
        \texttt{MarkovDiaryGenerator} & Markov Diary Learner and Generator. \\
        \hline
        \texttt{Gravity} & gravity model of human migration \cite{zipf1946p,barbosa2018human} \\
        \hline
        \texttt{Radiation} & radiation model for human migration \cite{simini2012universal} \\
    \end{tabular}
    \caption{Generative models implemented in \emph{scikit-mobility}.}
    \label{tab:genmodels}
\end{table}

\section{Privacy Risk Assessment}
\label{sec:privacy}
Mobility data is sensitive since the movements of individuals can reveal confidential personal information or allow the re-identification of individuals in a database, creating serious privacy risks \cite{demontoye2013unique, demontoye2018privacy}. 
Indeed the General Data Protection Regulation (GDPR) explicitly imposes on data controllers an assessment of the impact of data protection for the riskiest data analyses. For this reason, \emph{scikit-mobility} provides scientists in the field of mobility analysis with tools to estimate the privacy risk associated with the analysis of a given data set.

In the literature, privacy risk assessment relies on the concept of re-identification of a moving object in a database through an attack by a malicious adversary \cite{pellungrini2017data}. A common framework for privacy risk assessment \cite{PratesiMTGPY18} assumes that during the attack a malicious adversary acquires, in some way, the access to an anonymized mobility data set, i.e., a mobility data set in which the moving object associated with a trajectory is not known. Moreover, it is assumed that the malicious adversary acquires, in some way, information about the trajectory (or a portion of it) of an individual represented in the acquired data set. Based on this information, the risk of re-identification of that individual is computed estimating how unique that individual's mobility data are with respect to the mobility data of the other individuals represented in the acquired data set \cite{pellungrini2017data}.

\emph{scikit-mobility} provides several attack models, each implemented as a python class. For example in a location attack model, implemented in the \texttt{LocationAttack} class, the malicious adversary knows a certain number of locations
visited by an individual, but they do not know the temporal order of the visits \cite{pellungrini2017data}.
To instantiate a \texttt{LocationAttack} object we can run the following code:

\begin{verbatim}
Python> import skmob
Python> from skmob.privacy import attacks
Python> at = attacks.LocationAttack(knowledge_length=2)
\end{verbatim}

The argument \texttt{knowledge\_length} specifies how many locations the malicious adversary knows of each object's movement. The re-identification risk is computed based on the worst possible combination of \texttt{knowledge\_length }locations out of all possible combinations of locations. 

To assess the re-identification risk associated with a mobility data set, represented as a \texttt{TrajDataFrame}, we specify it as input to the \texttt{assess\_risk} method, which returns a \emph{pandas} DataFrame that contains the \texttt{uid} of each object in the \texttt{TrajDataFrame} and the associated re-identification risk as the column \texttt{risk} (type: float, range: $[0, 1]$ where 0 indicates minimum risk and 1 maximum risk).

\begin{verbatim}
Python> tdf = TrajDataFrame.from_file(filename="privacy_sample.csv")
Python> tdf_risk = at.assess_risk(tdf)
Python> print(tdf_risk.head())
\end{verbatim}
\begin{verbatim}
  uid        risk 
0   1    0.333333
1   2    0.500000
2   3    0.333333
3   4    0.333333   
4   5    0.250000   
\end{verbatim}

Since risk assessment may be time-consuming for more massive datasets, \emph{scikit-mobility} provides the option to focus only on a subset of the objects with the argument \texttt{targets}. For example, in the following code, we compute the re-identification risk for the object with \texttt{uid} 1 and 2 only:

\begin{verbatim}
Python> tdf_risk = at.assess_risk(tdf, targets=[1,2])
Python> print(tdf_risk)
\end{verbatim}
\begin{verbatim}
  uid        risk 
0   1    0.333333
1   2    0.500000
\end{verbatim}

During the computation, not necessarily all combinations of locations are evaluated when assessing the re-identification risk of a moving object: when the combination with maximum re-identification risk (e.g., risk 1) is found for a moving object, all the other combinations are not computed, so as to make the computation faster. However, if the user wants that all combinations are computed anyway, they can set the argument \texttt{force\_instances} (type: boolean, default: \texttt{False}) to \texttt{True}:

\begin{verbatim}
Python> tdf_risk = at.assess_risk(tdf, targets=[2], force_instances=True)
Python> print(tdf_risk)
\end{verbatim}

\begin{verbatim}
          lat        lng            datetime  uid  instance  instance_elem  \
0   43.843014  10.507994 2011-02-03 08:34:04    2         1              1   
1   43.708530  10.403600 2011-02-03 09:34:04    2         1              2   
2   43.843014  10.507994 2011-02-03 08:34:04    2         2              1   
3   43.843014  10.507994 2011-02-04 10:34:04    2         2              2   
4   43.843014  10.507994 2011-02-03 08:34:04    2         3              1   
5   43.544270  10.326150 2011-02-04 11:34:04    2         3              2   
6   43.708530  10.403600 2011-02-03 09:34:04    2         4              1   
7   43.843014  10.507994 2011-02-04 10:34:04    2         4              2   
8   43.708530  10.403600 2011-02-03 09:34:04    2         5              1   
9   43.544270  10.326150 2011-02-04 11:34:04    2         5              2   
10  43.843014  10.507994 2011-02-04 10:34:04    2         6              1   
11  43.544270  10.326150 2011-02-04 11:34:04    2         6              2   

    prob  
0   0.25  
1   0.25  
2   0.50  
3   0.50  
4   0.25  
5   0.25  
6   0.25  
7   0.25  
8   0.25  
9   0.25  
10  0.25  
11  0.25
\end{verbatim}

The result is a \emph{pandas} DataFrame that contains and a reference number of each combination under the attribute \texttt{instance} and, for each instance, the \texttt{risk} and each of the locations comprising that instance indicated by the attribute \texttt{instance\_elem}.
In Table \ref{tab:attacks}, we list the privacy attacks available in the library.

\begin{table}[]
    \centering
    \begin{tabular}{c|p{10cm}}
        \bf attack model & \bf assumed background knowledge \\
        \hline
        \texttt{LocationAttack} & locations visited by an object\\
        \hline
        \texttt{LocationSequenceAttack} & temporal sequence of locations visited by an object \\
        \hline
        \texttt{LocationTimeAttack} & locations visited by an object and the time of visit \\
        \hline
        \texttt{UniqueLocationAttack} & unique locations visited by an object, disregarding repeated visits to the same location \\
        \hline
        \texttt{LocationFrequencyAttack} & unique locations visited by an object and frequency of visitation \\
        \hline
        \texttt{LocationProbabilityAttack} & unique locations visited by an object and probability of visiting each location \\
        \hline
        \texttt{LocationProportionAttack} & unique locations visited by an object and relative proportion of the frequencies of visit \\
        \hline
        \texttt{HomeWorkAttack} & two most visited locations by an object \\
    \end{tabular}
    \caption{List of privacy attacks implemented in \emph{scikit-mobility}.}
    \label{tab:attacks}
\end{table}

%In its current version, \emph{scikit-mobility} implements eight different attack models, briefly summarized in Table \ref{tab:attacks} 

%% -- Illustrations ------------------------------------------------------------

%% - Virtually all JSS manuscripts list source code along with the generated
%%   output. The style files provide dedicated environments for this.
%% - In R, the environments {Sinput} and {Soutput} - as produced by Sweave() or
%%   or knitr using the render_sweave() hook - are used (without the need to
%%   load Sweave.sty).
%% - Equivalently, {CodeInput} and {CodeOutput} can be used.
%% - The code input should use "the usual" command prompt in the respective
%%   software system.
%% - For R code, the prompt "R> " should be used with "+  " as the
%%   continuation prompt.
%% - Comments within the code chunks should be avoided - these should be made
%%   within the regular LaTeX text.

%% -- Summary/conclusions/discussion -------------------------------------------

\section{Conclusion and Future Developments} \label{sec:summary}
In this paper, we presented \emph{scikit-mobility}, a new python library for the analysis, generation, and privacy risk assessment of mobility data. \emph{scikit-mobility} allows the user to manage two basic types of mobility data -- trajectories and fluxes -- and it provides several modules, each dedicated to a specific aspect of mobility data analysis.

\emph{scikit-mobility} has the advantage of providing, in a single environment, functions to deal with different aspects of mobility analysis, such as data preprocessing and cleaning, computation of mobility metrics, generation of synthetic trajectories and flows, and the assessment of privacy risk.
The current version of the library has some limitations too. For example, since \ee{pandas} \texttt{DataFrame}s must be fully loaded in memory, the size of the mobility data set that can be analyzed is limited by the capacity of the memory of the user's machine. Moreover, the library is currently designed to work with the latitude and longitude reference system only; it could be easily adapted to work with any reference system.

We imagine two future directions for the development of \emph{scikit-mobility}. On one side, we plan to add more modules to cover a more extensive range of aspects regarding mobility data analysis. 
For example, we plan to include algorithms for predicting the next location visited by an individual \cite{luca2020deep, wu2018location}. 
We will also consider including a module for performing map matching, i.e., assigning the points of a trajectory to the street network, and a module to compute the similarity between trajectories.

On the other hand, we plan to improve the library from a computational point of view. Although in its current version \emph{scikit-mobility} is easy to use and it is rather efficient on mobility data sets in the order of gigabytes, it is not scalable to massive mobility data in the order of terabytes. 
Since new python libraries similar to \emph{pandas} but more computationally efficient are being developed every year (e.g., \emph{dask}, \cite{rocklin2015dask}), we plan to re-implement crucial functions in \emph{scikit-mobility} so that they can exploit the computational efficiency of these libraries. This aspect, which is not crucial now, will become so when the library will be largely adopted by the scientific community.

\section{Existing tools}
\label{sec:related}
In this section, we briefly describe some of the existing libraries and tools that provide functionalities for movement data management. Overall, to the best of our knowledge, none of the other packages is tailored explicitly for human mobility, and none of them includes functions for privacy risk assessment. 
In Table \ref{tab:comparison}, we give a summary of the packages and functionalities.

\begin{table}[]
    \centering
    \begin{tabular}{c|c|c|c|c|c|c}
         & \bf data type & \bf processing & \bf plotting & \bf measures & \bf models & \bf privacy \\
         \hline
         \emph{scikit-mobility} & many & yes & yes & yes & yes & yes\\
         \emph{bandicoot} & mobile phone
         & yes & - & yes & - & - \\
         \emph{movingpandas} & many & yes & yes & - & - & - \\
         \emph{spacetime} & many & yes & yes & - & - & -\\
         \emph{trajectories} & many & yes & yes & yes & - & - \\
         \emph{adehabitatLT} & many & yes & yes & - & - & - \\
         \emph{TrajDataMining} & many & yes & yes & - & - & - \\
    \end{tabular}
    \caption{Comparison of scikit-mobility with other libraries that cover similar aspects of spatio-temporal and mobility data.}
    \label{tab:comparison}
\end{table}

\subsection*{R}
A review of state of the art reveals that several libraries (more than 50) deal with trajectory data in \texttt{R} \cite{joo2020rpacks}.
In the following, we give a brief overview of the packages that are the closest in the scope to \emph{scikit-mobility}.

\subsubsection*{Spacetime}
The \emph{spacetime} package \cite{spacetime} provides methods and functionalities from two other \texttt{R} packages, \emph{sp}  \cite{sp} and \emph{xts}  \cite{xts}. 
Package \emph{sp}  deals with different spatial data such as polygons, shapes, lines, or points, while package \emph{xts} handles time and dates. \emph{spacetime} provides several functionalities for the handling of spatio-temporal data, such as interpolation and calculation of empirical orthogonal functions. For visualizing data, \emph{spacetime} relies on other \texttt{R} packages, for example \emph{maps} \cite{maps} is used to draw geographical maps.

\subsubsection*{Trajectories}
The package \emph{trajectories} \cite{trajR} builds on the foundation of \emph{spacetime} providing a wider set of tools for managing non-domain specific trajectory data. It allows for the handling of single tracks of movement for each agent, and plotting and simulation of trajectories of different nature. 
It also provides model fitting for studying the behavior of individual tracks.

\subsubsection*{Adehabitat}
The packages under the \emph{adehabitat} family \cite{adehabitatORIGINAL} cover methods and functions to manage animal movement and habitat selection. 
Given the large number of available functions, the original package has been split into multiple smaller packages dealing with different aspects of animal movement: \emph{adehabitatHR} deals with home-range analysis, \emph{adehabitatHM} deals with habitat selection analysis, \emph{adehabitatLT} deals with animal trajectory analysis, and \emph{adehabitatMA} deals with maps. 
Many of the functions presented in these packages are specific for animal movement. 
\emph{adehabitatLT} \cite{Calenge2011AnalysisOA} is the most similar library to \emph{scikit-mobility}. 
However, \emph{adehabitatLT} deals mainly with trajectories sampled at regular time intervals, and it does not implement the individual and collective measures in \emph{scikit-mobility}, nor the individual and collective models of human mobility. 
\emph{scikit-mobility} is specifically designed to handle human mobility data, and therefore many of the models and methods provided by the \emph{adehabitat} packages cannot fully reproduce the same results.

\subsubsection*{TrajDataMining}
\ee{TrajDataMining} \cite{monteiro18Traj} provides some methods for trajectory data preparation, such as filtering, compression and clustering. It also provides some pattern recognition tools to extract recurrent movement behaviors from the trajectories. However, it does not implement generative models, one of the key features of \emph{scikit-mobility}, nor advanced plotting functionalities.

\subsection*{Python}
As for python, some libraries have been proposed to manage and manipulate mobility data. In this section, we revise the libraries that are the most similar in their purpose to what we propose in this paper, highlighting the differences between them and \emph{scikit-mobility}.

\subsubsection*{Bandicoot}
\emph{bandicoot} \cite{demontjoye2016bandicoot} is a python library for the analysis of mobile phone metadata that provides the users with functions to compute features related to mobile phone usage. These features are grouped into three categories: \emph{(i)} individual features describe an individual's mobile phone usage and interactions with their contacts;
\emph{(ii)} spatial features describe an individual's mobility patterns; 
\emph{(iii)} social network features describe an individual's social network.

The principal limit of \emph{bandicoot} is that it is specifically designed for managing a specific data type, namely mobile phone data. This design choice makes \emph{bandicoot} unsuitable for the analysis of movements that cannot be captured by mobile phone data, such as car travels, movements of animals, or boat trips. In contrast, \emph{scikit-mobility} gives the user the possibility to deal with a diverse set of mobility data sources (e.g., GPS data, social media data, mobile phone data) and covers a much complete set of standard mobility measures. Moreover, \emph{scikit-mobility} provides a module dedicated to the privacy risk assessment of any mobility data source, a module to create interactive geographic plots, and a module dedicated to generative models of individual and collective mobility, all features that are completely absent in \emph{bandicoot}.

\subsubsection*{Moving Pandas}
\emph{movingpandas} \cite{graser2019movingpandas} is an extension to the Python data analysis library \emph{pandas} \cite{pandas} and its spatial extension \emph{geopandas} \cite{geopandas} to add functionality for dealing with trajectory data. In \emph{movingpandas}, a trajectory is a time-ordered series of geometries. These geometries and associated attributes are stored in a \texttt{GeoDataFrame}, a data structure provided by the \emph{geopandas} library. The main advantage of \emph{movingpandas} is that, being based on \emph{geopandas}, it allows the user to perform several operations on trajectories, such as clipping them with polygons and computing intersections with polygons. However, since it is focused on the concept of trajectory, \emph{movingpandas} does not implement any features that are specific of mobility analysis, such as statistical laws of mobility, generative models, standard pre-processing functions, and methods to assess privacy risk in mobility data. 

%% -- Optional special unnumbered sections -------------------------------------

\section*{Acknowledgments}
Luca Pappalardo and Roberto Pellungrini have been supported by EU project SoBigData++ RI grant \#871042, EU project Track\&Know H2020 grant \#780754 
and by EPSRC First Grant EP/P012906/1. 
We thank Anita Graser for the useful discussions, and Massimiliano Luca, Giuliano Cornacchia, and Michele Ferretti for their contribution to the development of the library.

\section*{Author contributions}
L.P. developed modules, performed experiments, developed code examples, made the documentation, and structured the paper.
F.S. performed experiments, tested the code and developed modules. 
G.B. performed the code, the system design and developed modules. 
R.P. performed experiments and developed modules. 
All the authors contributed to writing of the manuscript.

%% -- Bibliography -------------------------------------------------------------
%% - References need to be provided in a .bib BibTeX database.
%% - All references should be made with \cite, \citet, \cite, \citealp etc.
%%   (and never hard-coded). See the FAQ for details.
%% - JSS-specific markup (\proglang, \pkg, \texttt) should be used in the .bib.
%% - Titles in the .bib should be in title case.
%% - DOIs should be included where available.
\bibliographystyle{apalike} 
\bibliography{refs}

\end{document}